\documentclass[twocolumn,tighten]{aastex701}
 
\usepackage{amsmath}
\usepackage{siunitx}

\newcommand{\halpha}{H$\alpha$}

\newcommand{\henir}{{\ion{He}{1}}$~\lambda$10830}

\newcommand{\av}{A$_{\text{V}}$}
\newcommand{\mdot}{$\dot{\text{M}}$}
\newcommand{\Mdot}{{\rm \dot{{M}}}}

\newcommand{\msun}{\rm M_{\sun}}

\newcommand{\rstar}{\rm R_{\star}}

\newcommand{\msunyr}{\rm{M_{\sun} \, yr^{-1}}}

\newcommand{\kms}{\rm \, km \, s^{-1}}

\newcommand{\ri}{R$_{\rm i}$}
\newcommand{\rw}{W$_{\rm r}$}
\newcommand{\tmax}{T$_{\rm max}$}

\newcommand{\jwst}{\emph{JWST}}
\newcommand{\spitzer}{\emph{Spitzer}}

\newcommand{\tess}{\emph{TESS}}
\newcommand{\thestar}{CVSO~1942}
 
\newcommand{\newarcsec}[1]{\ang[angle-symbol-over-decimal]{;;#1}}

\received{2025 December 1}
\revised{2026 January 28}
\accepted{2026 February 8}
 
\shorttitle{CVSO~1942 Variability}
\shortauthors{Thanathibodee et al.}

\begin{document}
 
\title{Indications of Rapid Dust Formation in the Inner Region of a Protoplanetary Disk}
 
\correspondingauthor{Thanawuth Thanathibodee}
\email{thanawuth.t@chula.ac.th}
 
\author[0000-0003-4507-1710]{Thanawuth Thanathibodee}
\affiliation{Department of Physics, Faculty of Science, Chulalongkorn University, 254 Phayathai Road, Pathumwan, Bangkok 10330, Thailand}
\email{thanawuth.t@chula.ac.th}
 
\author[0000-0001-9227-5949]{Catherine Espaillat}
\affiliation{Department of Astronomy, Boston University, 725 Commonwealth Avenue, Boston, MA 02215, USA}
\affiliation{Institute for Astrophysical Research, Boston University, 725 Commonwealth Avenue, Boston, MA 02215, USA}
\email{cce@bu.edu}
 
\author[0000-0002-3950-5386]{Nuria Calvet}
\affiliation{Department of Astronomy, University of Michigan, 1085 South University Ave., Ann Arbor, MI 48109, USA}
\email{ncalvet@umich.edu}
 
\author[0000-0003-3616-6822]{Zhaohuan Zhu}
\affiliation{Department of Physics and Astronomy, University of Nevada, Las Vegas, 4505 S. Maryland Pkwy, Las Vegas, NV 89154, USA}
\affiliation{Nevada Center for Astrophysics, University of Nevada, Las Vegas, 4505 S. Maryland Pkwy, Las Vegas, NV 89154, USA}
\email{zhaohuan.zhu@unlv.edu}
 
\author[0009-0009-7491-7720]{Julalak Nammanee}
\affiliation{Department of Physics, Faculty of Science, Chulalongkorn University, 254 Phayathai Road, Pathumwan, Bangkok 10330, Thailand}
\email{julalaknammanee@gmail.com}
 
\author[0000-0001-9301-6252]{Caeley Pittman}
\affiliation{Department of Astronomy, Boston University, 725 Commonwealth Avenue, Boston, MA 02215, USA}
\affiliation{Institute for Astrophysical Research, Boston University, 725 Commonwealth Avenue, Boston, MA 02215, USA}
\email{cpittman@bu.edu}
 
\author[0009-0005-4517-4463]{M\'aire Volz}
\affil{Department of Astronomy, Boston University, 725 Commonwealth Avenue, Boston, MA 02215, USA}
\affil{Institute for Astrophysical Research, Boston University, 725 Commonwealth Avenue, Boston, MA 02215, USA}
\email{mvolz@bu.edu}

\begin{abstract}
We \added{report} a significant increase in mid-infrared emission $\leq10\,\mu$m in a transitional disk. The 2024 JWST/MIRI observation of the disk around CVSO~1942 reveals flux increase by a factor of two at $\lambda\leq10\,\mu$m, compared to the near photospheric flux level observed with {\spitzer}/IRS in 2005. No significant change in flux at $\gtrsim15\mu$m is detected in the spectra. Comparing the MIRI/MRS spectrum and NEOWISE photometry, we found that this $\leq10\,\mu$m flux increase occurs on a timescale of 2 weeks and is consistent with the presence of warm (1,400\,K), optically thick, large ($\gtrsim1\,\mu$m) dust grains near the dust sublimation radius. We propose that this rapid dust appearance may indicate in situ dust formation, possibly from planetesimal collisions in the inner disk.
 
\end{abstract}


\section{Introduction} \label{sec:intro}
Variability is a hallmark of low-mass, pre-main sequence stars (i.e., T Tauri stars; TTS). Despite being a defining characteristic \citep{joy1945}, the physics behind TTS variability is not fully understood. This is especially true in the infrared, where the emission is dominated by the surrounding protoplanetary disk.
 
Studies of infrared variability \citep{cody2014,wolk2018,lee2024} were made possible by space-based observations, finding possible links between protoplanetary disk evolution stages and magnitudes of variability. Spectroscopic observations have provided insight into the causes of these variations, for example, identifying see-saw-type variability linked to changing inner disk structure \citep{espaillat2011}. Nevertheless, the causes of other types of variability, such as infrared bursts, identified in photometric studies \citep[e.g.,][]{lee2024} are less clear.

Recent {\jwst} observations and archival {\spitzer} observations have discovered many aspects of infrared variability \citep{espaillat2024,jang2024,xie2025}. Here, we present a study of infrared variability of {\thestar}, focusing on its disk properties as traced by mid-infrared spectroscopy using new and archival data. CVSO~1942 is a TTS of spectral type K6 (previously published as a K4 star and called FM~177 in \citealt{espaillat2012}) located in Cloud B of the Orion OB1 star-forming region \citep{briceno2019} \added{with A$_{V}=0.6$}. The star was identified as a low accretor based on a clear redshifted absorption feature on the {\henir} line profile \citep{thanathibodee2022} despite the weak {\halpha} emission. The low accretion rate of $\Mdot=(1.3\pm0.4)\times10^{-10}\,\msunyr$ \citep{thanathibodee2023} along with its transitional disk \citep{espaillat2012} suggests that the disk is near the last stages of its evolution. 
 
In Section 2, we describe the observations of the star and present our analysis in Section 3. We discuss the implications of the results in Section 4 and provide a summary in Section 5.
 
\begin{figure*}[t]
\epsscale{1.1}
\plotone{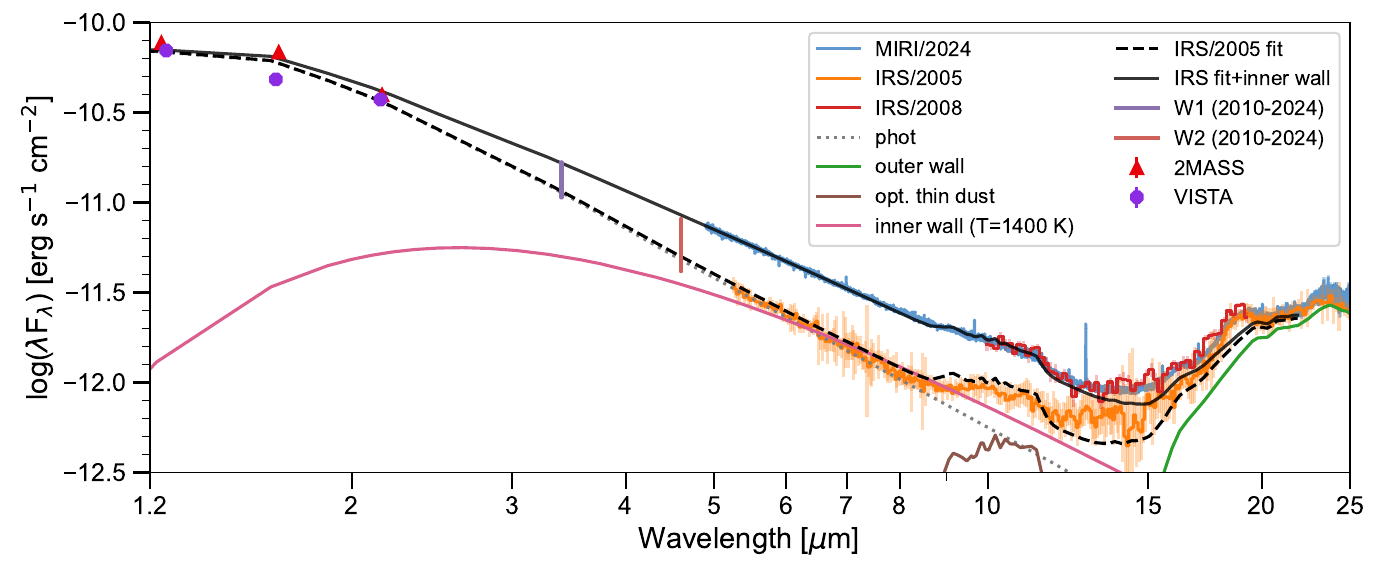}
\caption{Infrared Spectral Variability of CVSO~1942.
The blue, orange, and red lines are observations from JWST/MIRI (2024), {\spitzer}/IRS-LowRes (2005), and {\spitzer}/IRS-HiRes (2008), respectively. The best fit to the IRS-LowRes spectrum is shown as a black dashed line, which represents the combination of the photosphere (gray dotted line), the optically thin dust (brown solid line), and the outer wall (green solid line), as described in \citet{espaillat2012}. By adding a blackbody component with T=1400\,K (pink solid line) to the 2008 {\spitzer} model, we can fit the {\jwst}/MIRI observation. The purple and red vertical bars denote the range of observations with WISE W1 and W2, shown in Fig.~\ref{fig:wise}. 
\label{fig:sed}}
\end{figure*}
 
\section{Observations and Data Reduction} \label{sec:observation}
We obtained new data and gathered archival data to study the infrared variability of {\thestar}. New data were obtained with {\jwst} and WIYN, while archival data were obtained from previous observations with {\spitzer}, the Magellan Telescope, TESS, and WISE. We provide the details of the observations and data sources below.
 
\subsection{Infrared Data}
\subsubsection{JWST}
{\thestar} was observed on 2024 March 10 with the MIRI/MRS instrument on board {\jwst}, \added{covering $4.9-27$\,$\mu$m}, as part of the program GO-3983 (PI: T. Thanathibodee). The total exposure time is 666\,s per configuration with 4-point dithers. As the star is isolated, no dedicated background observation was obtained. We reduced the data using the procedure outlined in \citet{espaillat2023} with the MIRI pipeline version 1.20.1, setting the CRDS Context to \texttt{jwst\_1464.pmap}. We present the spectrum as the blue line in Figure~\ref{fig:sed}.

\subsubsection{Spitzer}
The star was observed twice by the {\spitzer} telescope during its operation. First, it was observed with the IRS instrument in low-resolution mode ($5.2-37$\,$\mu$m) on 2005 October 15 as part of Program ID 85 (PI: G. Rieke). We used the reduced spectrum from 
\citet{espaillat2012},
shown here as the orange line in Figure~\ref{fig:sed}.
 
Another IRS observation occurred on 2008 November 17 in the high-resolution mode as part of Program ID 40247 (PI: N. Calvet); the data were first presented in \citet{espaillat2013}. We downloaded the spectrum from CASSIS \citep{lebouteiller2015}, selecting the optimal extraction method, and re-binned the spectrum by 15 pixels. The spectrum covers $10-20$\,$\mu$m wavelength range and is presented as the red line in Figure~\ref{fig:sed}.
 
\begin{figure*}[t]
\epsscale{1.1}
\plotone{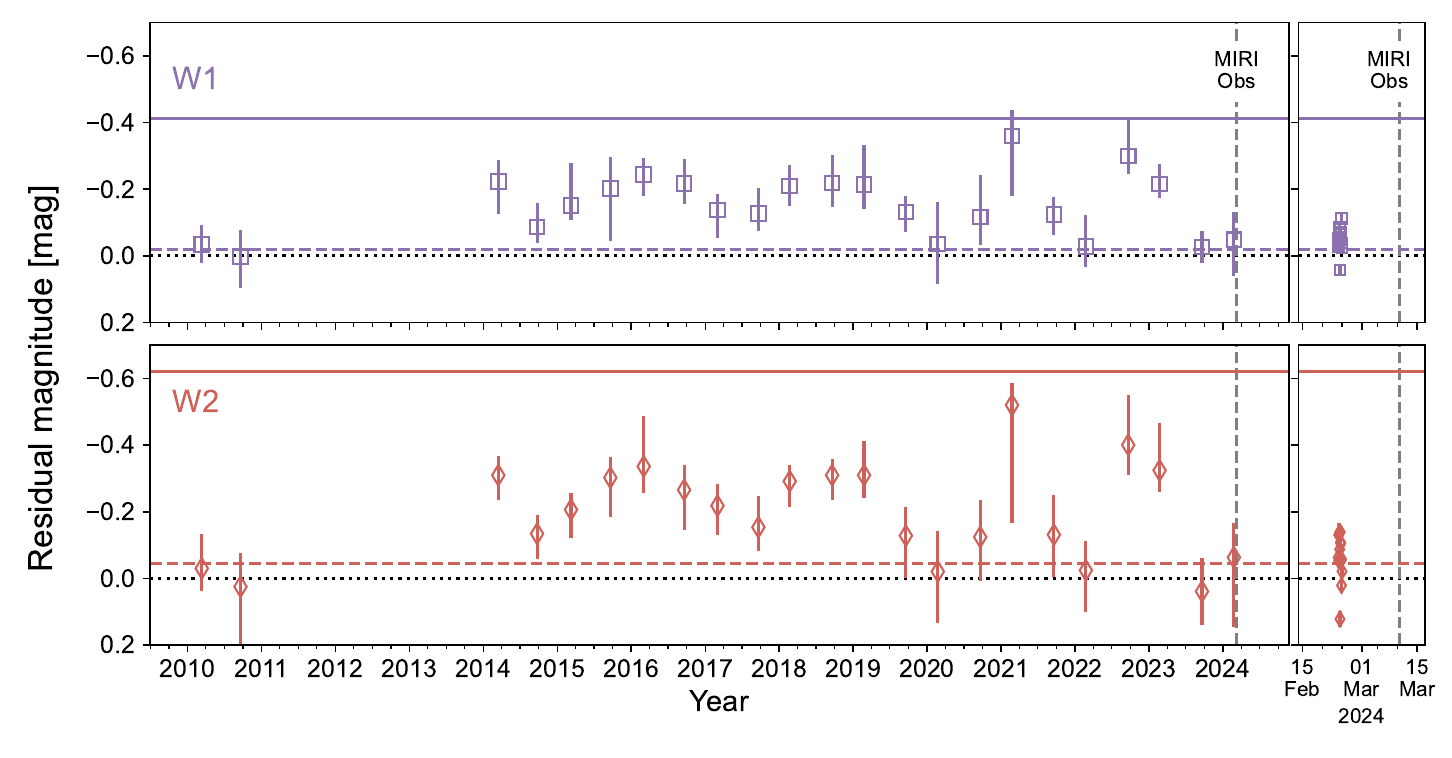}
\caption{\emph{Left:} The light curve of CVSO~1942 from the ALLWISE (2010-2011) and NEOWISE (2014-2024) survey. The photosphere has been subtracted. The horizontal lines indicate the levels extrapolated by the models that fit the 2008 {\spitzer} observation (dashed lines) and the 2024 {\jwst} observation (solid lines). The horizontal dotted lines show the photospheric level, and the vertical dashed line shows the date of the {\jwst} observation, which was 2 weeks after the last NEOWISE observation. The data points are binned to 7-days. \emph{Right:} The zoom-in, unbinned light curve of the star around the {\jwst} observation.
\label{fig:wise}}
\end{figure*}
 
\subsubsection{WISE}
For comparison with the {\jwst} and {\spitzer} data, we retrieved mid-infrared photometry in the WISE W1 \added{(3.4\,$\mu$m)} and W2 \added{(4.6\,$\mu$m)} bands from \added{IRSA}, covering the data from the cryogenic WISE mission \citep{wright2010}, the post-cryogenic NEOWISE program between 2010 and 2011 \citep{mainzer2011}, and the NEOWISE reactivation program for data 2014 onward \citep{mainzer2014}. The latest photometry was observed on 2024 February 25, which was 14 days prior to the {\jwst} observation. The photometry is shown in Figure~\ref{fig:wise}.
 
\subsection{Optical Data}
 
\begin{figure*}[t]
\epsscale{1.1}
\plotone{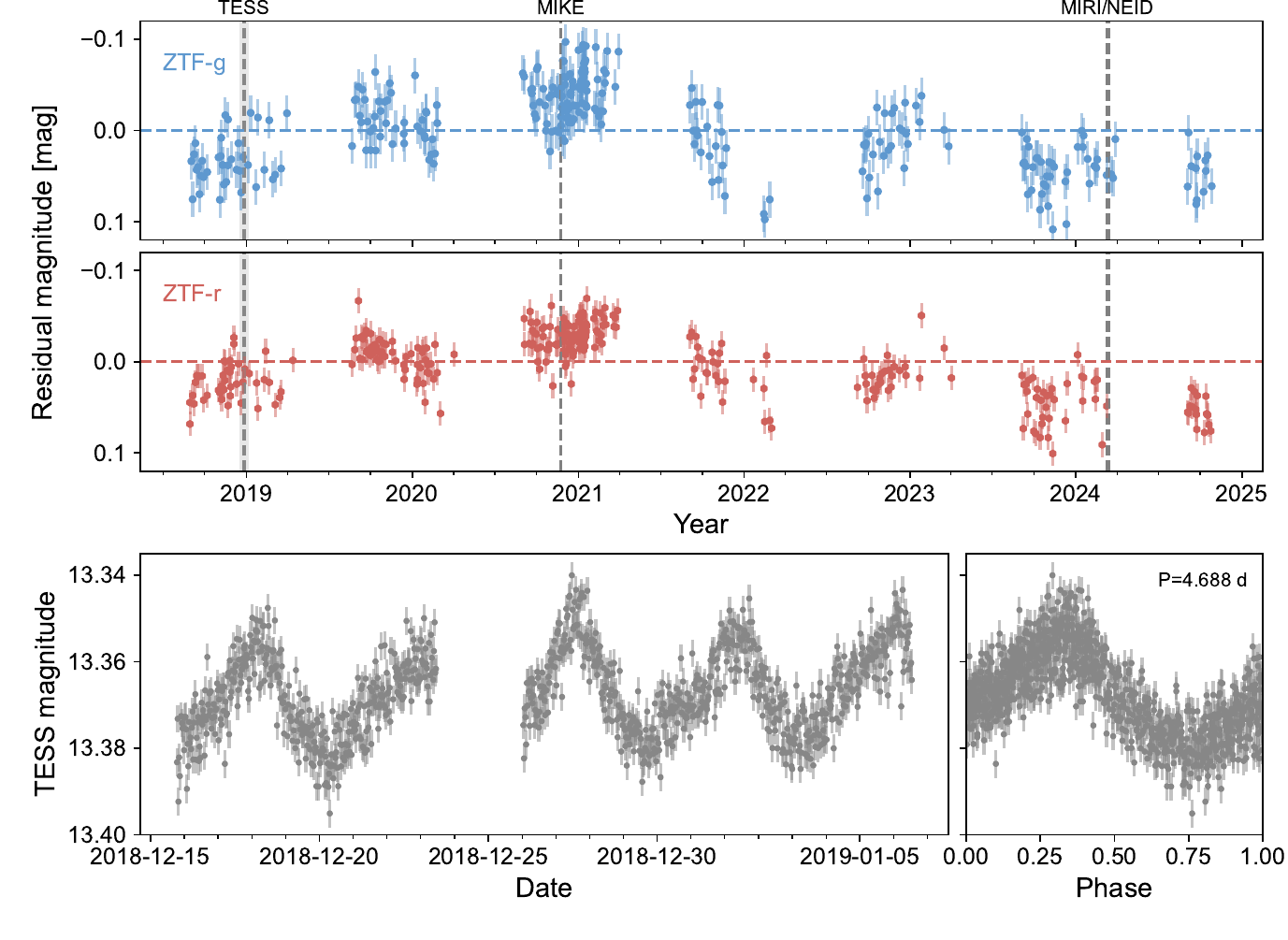}
\caption{Optical light curves of CVSO~1942.
\emph{Top/middle:} The photometry from ZTF DR23 in the g and r bands. The data points are subtracted by the median magnitudes (shown in dashed lines). Vertical lines mark the observation dates of {\tess}, Magellan/MIKE (2020), {\jwst}/MIRI, and WIYN/NEID.
\emph{Bottom:} The {\tess} light curve. The optical light curve is periodic with a period of 4.688 days.
\label{fig:optical}}
\end{figure*}
 
\subsubsection{ZTF}
We downloaded the optical photometry of {\thestar} obtained with the Zwicky Transient Facility (ZTF) via NASA/IPAC Infrared Science Archive. As of DR23, there are 309 data points in the ZTF-r band and 277 data points in the ZTF-z band for which \texttt{catflags=0}. These data cover the observation period from 2018 August 29 to 2024 October 26. We show the data in the top two panels of Figure~\ref{fig:optical}.
 
\subsubsection{TESS}
Broad-band optical data of {\thestar} were obtained by the Transiting Exoplanet Survey Satellite (TESS) in Sector 6, which covers the period between 2018 December 12 to 2019 January 6. The photometry was obtained using TESSExtractor \citep{serna2021,brasseur2019} and is shown in the bottom panel of Figure~\ref{fig:optical}. The star is relatively isolated, with Crowding Ratio = 0, so the photometric results should be reliable. 
 
\subsubsection{WIYN}
We obtained a new \added{spectroscopic} observation of {\thestar} using the NEID instrument at the 3.5m WIYN telescope, Kitt Peak National Observatory, as part of program 2024A-408522 (PI: C. Pittman). The data were obtained three days after {\jwst}, on 2024 March 13, in the high efficiency (HE) mode with the \newarcsec{1.5} fiber, providing the resolution R$=60,000$. The data were reduced with the NEID reduction pipeline and the two 1D spectra were combined with a custom code.
 
\subsubsection{Magellan}
{\thestar} was observed on 2009 January 19 and 2020 November 23 with the MIKE instrument at the Magellan Clay Telescope at the Las Campanas Observatory in Chile. Using the same configuration of \newarcsec{0.7} slit and $2\times2$ binning, the resulting spectra have the resolution R$=32,500$. The data were reduced using the MIKE pipeline. These data have been previously published by \citet{espaillat2012} and \citet{thanathibodee2023}, and were included here for re-analysis and comparison.

\section{Analysis and Results} \label{sec:results}
\subsection{Infrared Spectral Variability} \label{ssec:ir_spec_var}
We present the spectral energy distribution  of the system from $1\,\mu$m to $25\,\mu$m in Figure~\ref{fig:sed}, where we include magnitudes from 2MASS \citep{skrutskie2006} \added{and VISTA \citep{mcmahon2013}}. The photosphere (dotted line) was calculated using colors for K6 young stars from \citet{pecaut2013}, scaled to the 2MASS J magnitude. In 2005 (Spitzer/IRS, orange line), the spectrum at short wavelengths ($\lesssim8\,\micron$) is mostly consistent with the star's photosphere while the other IRS spectrum (red) and the {\jwst} spectrum (blue) observed at later times show significant excess over the photosphere at this wavelength range. Interestingly, at wavelengths longer than 15\,{\micron}, the {\jwst}/MIRI flux is similar to all prior observations. While MIR continuum variability has been observed previously \citep{espaillat2011,jang2024,xie2025,smith2025}, this observation represents \added{one of} the most significant {\it increases} in MIR emission, by a factor of two, seen to date. Evidence of a previous increase in flux may be inferred from the 2008 IRS observations, where the $10-20\,\mu$m spectrum is comparable to that observed in 2024.
 
We plot, in Figure~\ref{fig:sed}, the results of the disk modeling by \citet{espaillat2012}, in which the authors fit the 2005 IRS data with the D'Alessio Irradiated Accretion Disk model \citep{dalessio2006}. At this time, the observations can be explained by an outer disk truncated at 49 au and an inner cavity filled with optically thin dust, as shown by the total model (black dashed line). The model components include the stellar photosphere, the optically thin inner disk (brown solid line), and the outer disk wall (green solid line).

With the increase in flux at wavelengths $\leq10\,\mu$m, the 2024 {\jwst} observation requires an extra emitting component in the system. 
\added{We fit the excess with a blackbody varying the temperature from $1,000-2,000$\, K and found that it is consistent with $1,200-1,700$\, K. We adopted $1,400$\,K as the representative temperature of the dust, consistent with the emission arising from the disk sublimation radius \citep{isella2005,dalessio2006}.}
The black solid line is the sum of the 2005 IRS fit and the 1,400\,K blackbody, and it is comparable to the {\jwst} observation at all wavelengths, suggesting that the difference between the two epochs can be accounted for by a single-temperature blackbody representing an inner optically thick region.

\subsection{Infrared Photometric Variability} \label{ssec:ir_phot_var}
We compare the extrapolated flux of the {\spitzer} and {\jwst} spectra with the WISE W1 and W2 flux observed over the 15 year period. Since the spectra do not extend below $\sim5\,\micron$, we assume that the flux of the system at the wavelengths of the W1 and W2 bands can be determined solely from the disk model that best fits the mid-infrared data. In  Figure~\ref{fig:sed}, we plot the full ranges of the W1 and W2 fluxes as observed since 2010 as vertical lines. As expected, the upper ranges are close to the level expected at the {\jwst} epoch, whereas the lower levels are slightly below those expected at the time of the 2005 {\spitzer} observation.
 
To investigate the long-term variability in the W1 and W2 bands, we plot the light curve of the system in Figure~\ref{fig:wise}. The data points are binned by $\sim7$ days and the error bars indicate the full range within the bin plus measurement uncertainties. The magnitudes have been subtracted by the {\av}-corrected, scaled, photospheric magnitude determined from Table~6 of \citet{pecaut2013}. For comparison, we also plot the magnitudes of the system during the {\jwst} epoch (solid lines) and the 2005 {\spitzer} epoch (dashed lines). Based on the light curve, it is evident that the flux levels of the last WISE observations are consistent with the near-photospheric levels observed in 2005 with {\spitzer}, whereas the fluxes at the {\jwst} epoch are among the highest levels of the past 15 years. Notably, the majority of the photometric points are closer to the photosphere than the {\jwst} level.
 
\added{Following the analysis in Section~\ref{ssec:ir_spec_var}, we determine the blackbody temperature consistent with the W1-W2 color excess on a 1-day basis. Within the uncertainty, we found no correlation between the brightness and the temperature, and that most data points are consistent with $1,200-1,700$\,K found above.}
 
We note that the last photometric point in 2024 was two weeks before the {\jwst} observation. That is, the change in flux from the photospheric level to the {\jwst} level occurs on a short timescale of $\lesssim$2 weeks.
 
\subsection{Optical Photometric Variability} \label{ssec:opt_phot_var}
We include optical light curves from the ZTF and {\tess} to connect the stellar and disk variability in Figure~\ref{fig:optical}. The top two panels show six-year observations in the ZTF g and r bands. For each band, we subtract the magnitudes by their medians. There is no change larger than $\sim0.1$ mag within a given observing season, and none more than $\sim0.2$\,~mag overall. No evidence of a flaring or dipping event is found.
 
The {\tess} spacecraft observed the star in 2019 during which its ZTF magnitudes were slightly fainter than the 6-year average. The lower panel of Figure~\ref{fig:optical} shows the light curve (left) with clear periodicity. Using \texttt{TESSExtractor}, we find a 4.688-day period, consistent with typical rotation periods of T Tauri stars \citep{karim2016,serna2021}. The phase-folded light curve is shown on the right. The ZTF light curve also covers the epoch of {\jwst} observation.

\begin{figure*}[t]
\epsscale{1.1}
\plotone{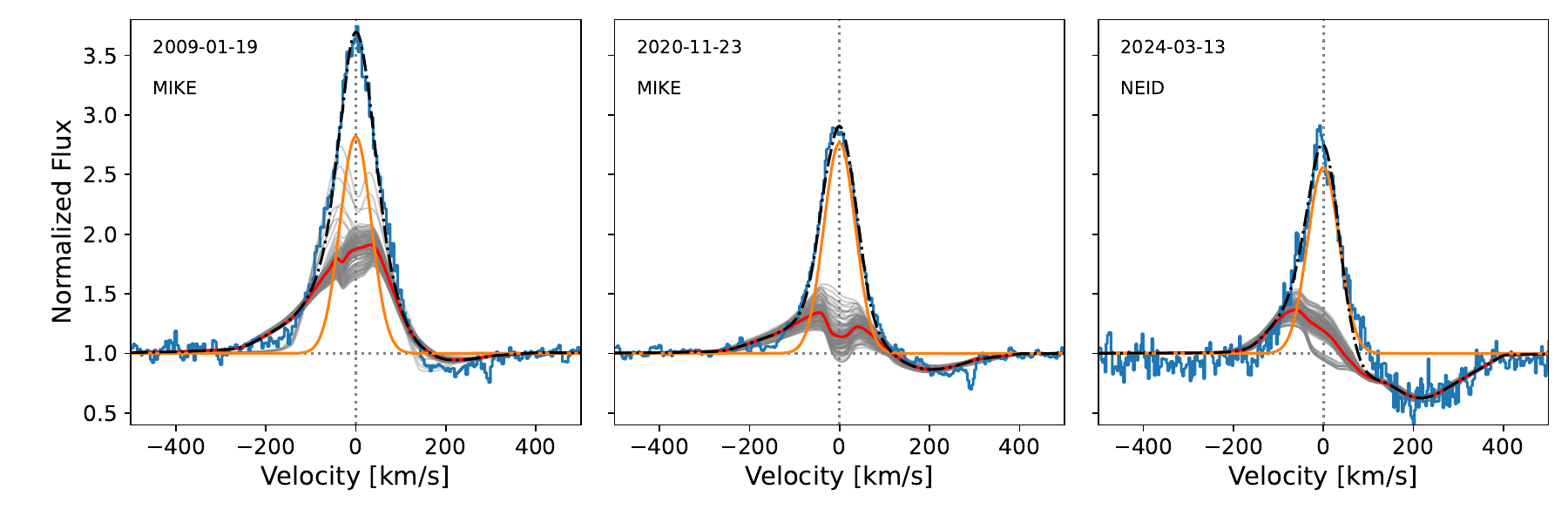}
\caption{{\halpha} profiles of CVSO~1942 across 15 years. The blue lines represent the observations, the red lines represent the average magnetospheric model, and the gray lines represent the top 100 best-fit models. The orange lines are the average chromospheric models, and the black lines are the average total models.
\label{fig:halpha}}
\end{figure*}
 
\subsection{Accretion Properties} \label{ssec:accretion}
 
We investigate the changes in the mass accretion rate from the disk onto the star across the years, by modeling {\halpha} profiles using accretion flow models of \citet{muzerolle2001,muzerolle1998a,hartmann1994}. This model adopts the magnetospheric accretion paradigm \citep[see][for a review]{hartmann2016}, assuming axisymmetric, dipolar accretion flows. The details of the code were described by the original papers. We create a large grid of 607,200 models varying the mass accretion rate ($5\times10^{-11}$ to $5\times10^{-10}\,\msunyr$), the disk truncation radius \ri\ (1.6 to 5.6 R$_{\star}$), the equatorial width of the flow \rw\ (0.2 to 2.0 R$_{\star}$), the maximum flow temperature \tmax\ (11,000 to 14,500 K), and inclinations of the system ($10^\circ$ to $80^\circ$).
 
We fit the observations following the procedure in \citet{thanathibodee2023}, in which a chromospheric contribution at the line center is assumed to be a Gaussian function with amplitude $A_c$ and standard deviation $\sigma_c$. We show the top 100 best fits in Figure~\ref{fig:halpha} and the results in Table~\ref{tab:model_results_acc}, calculated from more than $\sim5,000$ best fit models with the likelihood $\geq0.5$. The accretion parameters remain largely unchanged, with mass accretion rates within a factor of two across 15 years.
 
\begin{deluxetable*}{ccccccccc}[t!]
\tablecaption{Results of the Magnetospheric Accretion Model \label{tab:model_results_acc}}
\tablehead{
\colhead{Obs. Date} &
\colhead{Instrument} &
\colhead{\mdot} &
\colhead{\ri} &
\colhead{\rw} &
\colhead{\tmax} &
\colhead{$i$} &
\colhead{$A_c$} &
\colhead{$\sigma_c$} \\
\colhead{(UT)} &
\colhead{} &
\colhead{($10^{-10}\,\msunyr$)} &
\colhead{(R$_{\star}$)} &
\colhead{(R$_{\star}$)} &
\colhead{($10^3\,$K)} &
\colhead{(deg)} &
\colhead{} &
\colhead{(km\,s$^{-1}$)}
}
\startdata
2009-10-19 & MIKE  & 2.3$\pm$1.3 & 2.0$\pm$0.7 & 0.5$\pm$0.2 &  12.7$\pm$1.0 & 61$\pm$21 & 1.8$\pm$0.4 & 33.0$\pm$4.4 \\
2020-11-23 & MIKE  & 1.8$\pm$1.0 & 2.5$\pm$1.0 & 0.7$\pm$0.4 &  12.5$\pm$1.0 & 59$\pm$21 & 1.7$\pm$0.4 & 32.9$\pm$3.3 \\
2024-03-13 & NEID  & 3.0$\pm$1.1 & 2.3$\pm$0.5 & 1.6$\pm$0.3 &  12.8$\pm$1.0 & 52$\pm$10 & 1.7$\pm$0.3 & 34.6$\pm$1.4 \\
\enddata
\end{deluxetable*}

\section{Discussion} \label{sec:discussion}
 
We \added{report} a significant increase in mid-infrared flux from a protoplanetary disk by comparing the new {\jwst}/MIRI observation of {\thestar} in 2024 with archival {\spitzer}/IRS observations obtained in 2005 and 2008. We propose that the 1400\,K blackbody excess in the MIRI spectrum is due to optically thick dust with large grains. This is because the $10\,\micron$ silicate feature, which traces small grains, is unchanged from the 2005 IRS spectrum \citep{espaillat2012}. Furthermore, this dust emission has a variability timescale of $\sim$weeks, based on the 2-week difference between the last WISE observation and the {\jwst} observation.
 
We calculate the emitting area of the blackbody as
\begin{equation}
    A = d^2\Omega=d^2\left(\frac{F_{\nu, diff}}{B_{\nu}}\right)
\end{equation}
where $d=404$\,pc is the Gaia distance and the best-fit solid angle $\Omega=\frac{F_{\nu, diff}}{B_{\nu}}=1.2\times10^{-19}$\,Sr. We found the emitting area $A=1.93\times10^{23}$\,cm$^2$ at the dust sublimation radius\footnote{$R_d$ is calculated from the equilibrium temperature assuming dust grains behave like a blackbody.} $R_d=\left(\frac{T_{\star,eff}}{T_{dust}}\right)^2R_{\star}=8.2\,R_{\star}$ with $T_{dust}=1400$\,K.
 
As the dust is optically thick ($\tau=\Sigma\kappa_d\gtrsim1$), we adopt the dust opacity at $6\,\micron$ of $\kappa_{d}\approx200$\,cm$^2$g$^{-1}$ (model B11, B11S with maximum grain size of 1\,mm; \citealt{birnstiel2018}) to calculate the minimum dust mass of
\begin{equation}
    M_{d} \gtrsim\frac{A}{\kappa_d} = 9.7\times10^{20}\,{\text g} = 4.9\times10^{-13}\,\msun,
\end{equation}
with the opacity being the main source of uncertainty, \added{and an additional 30\% from the range of possible dust temperatures}.
 
In comparison, the total dust mass in the 49\,au-wide optically thin region calculated by \citet{espaillat2012} using the 2005 {\spitzer}/IRS data is $2\times10^{-12}\,\msun$. This suggests that the increase in 2024 represents a significant enhancement in the dust surface density over a small area in less than 2 weeks. This dust mass is also slightly larger than the $(1.2\pm0.4)\times10^{-13}\,\msun$ dust mass expected to be accreted in two weeks, calculated from the contemporaneous mass accretion rate (Section~\ref{ssec:accretion}) and the standard dust-to-gas mass ratio of 0.01. Assuming a larger dust-to-gas mass ratio would suggest even higher dust enhancement, as the observed 1400\,K dust represents the material that has not been accreted onto the star.

\subsection{Rapid Dust Formation}
We propose that the excess emission shortward of $10\,\micron$ during the {\jwst} observation came from rapid, in situ dust formation near the dust sublimation radius. This is consistent with the seemingly stochastic nature of the increased emission in the light curve (Fig.~\ref{fig:wise}) as well as a short timescale for this to occur. This dust formation may be accompanied by variability at longer timescales, such as the seemingly periodic changes between 2014 and 2020. The minimum dust mass required to create this excess is larger than the mass loss through accretion during the same time period. If we consider the peaks in the WISE light curve in early 2021 and mid/late 2022 as dust formation events, the subsequent data points suggest that dust draining can take $\sim$months, consistent with the mass accretion rate. 
 
A similar scenario may have occurred during the 2008 {\spitzer} observation, in which a different amount of dust was produced, compared to the 2024 {\jwst} observation. However, since the dust is optically thick, different amount of dust may result in a similar flux level at wavelengths longer than 10\,$\mu$m. That is, the fact that the 2024 and 2008 flux levels are similar does not necessarily rule out a stochastic dust formation mechanism.
 
Several scenarios may be consistent with the rapid dust formation proposed here. For example, this could be the case of dust formation due to collisions of planetesimals, similar to the process by which debris disks form. While it is unclear whether such a process can initiate while stars still host primordial disks, this mechanism has been observed to produce mid-IR variability in young debris disk systems \citep[e.g.,][]{su2019} with timescales of $\sim$months to $\sim$years. If this is the case, this will be the first observational evidence of active transition from a primordial protoplanetary disk to a debris disk at the last stages of accretion, where the accretion rates are low. However, more observations are needed to confirm the proposed scenario.
 
Other dust formation processes that may be relevant to explain the observations include thermally driven dust formation, such as those proposed by \citet{kato2025}. While the timescale for that proposed process is generally $\sim$years, it may be shorter in some scenarios, such as those with low viscosity. Another possibility that can solve the timescale problem is a cavitating bubble due to radiation-condensation instability proposed by \citet{chiang2024}. Small thermal variations in the silicate gas result in small dust particles that can undergo runaway growth. While its $\sim$hours timescale is compatible with our observations, more modeling work is required to test whether this scenario can produce as much dust as we estimate here.
 
A more elucidating explanation will require monitoring this object at all timescales, from hours to months, to distinguish between these scenarios. More modeling work, including studying the evolution of dust generated by planetesimal collisions in a gaseous disk and modeling the spectrum of dust generated by thermal instability and condensation, would be required. In any case, {\thestar} provides an excellent opportunity to study the coupling of dust and gas evolution near the terrestrial planet forming site, warranting further investigation.

\subsection{Alternative Explanations}
In addition to the process described above, we consider several other explanations here.
 
\subsubsection{Accretion Burst or Flaring}
Accretion in young stars is found to be variable at all timescales \citep{wendeborn2024a,herczeg2023}. Some of the accretion luminosity may manifest in the infrared \citep{kospal2018,pittman2022}.  Bursting or flaring could also be observed in the infrared \citep[e.g.,][]{morales-calderon2011,park2021}. Therefore, it may be the case that accretion variability causes the increased flux at wavelengths $\lesssim10\,\micron$. However, \citet{xie2025} compared {\jwst} and {\spitzer} data of T~Cha and found that an accretion burst can cause the destruction of the inner disk wall, resulting in see-saw variability, which is not seen in our observations. Moreover, flares/bursts are found to be correlated with the optical and infrared wavelengths. Although we do not have the infrared light curve at the time of the {\jwst} observation, the optical light curve (Fig.~\ref{fig:optical}) shows no sign that the star is a burster. Lastly, bursters' spectra usually peak at short wavelengths (X-ray, UV; e.g., \citealt{hinton2022}), which does not explain the 1,400\,K temperature of the increased MIR emission. Therefore, we conclude that a bursting or flaring event is not responsible for the increased emission at the {\jwst} epoch of CVSO~1942.

\subsubsection{Long-lived Dust Clump}
The dust sublimation region in the inner disk is thought to be associated with various processes that result in dust accumulation, which can persist for a long time \citep[e.g.,][]{flock2017,ueda2019}. Observations with interferometers have also found evidence of dust clumps near the dust sublimation radius \citep{setterholm2025}. Can an occultation of such a disk feature explain the variable mid-IR observations of {\thestar}? 
 
Adopting the {\tess} period of 4.688\,d as the star's rotation period, and the $v\sin i = 7.99\,\kms$ \citep{thanathibodee2023}, the inclination of the star is $\sim30\deg$. At this low inclination, it is unlikely that a feature at the dust sublimation radius ($R_d=8.2\,R_{\star}$) would be occulted by the star or the magnetosphere ($\sim4\,\rstar$; Section~\ref{ssec:accretion}). Moreover, if such a feature exists, there should be $\sim50\%$ chance (or more, given the geometry) that the system will be observed in the bright state, which is not the case according to Figure~\ref{fig:wise}. Therefore, a long-lived dust clump near the dust sublimation radius is not consistent with the observations.

\subsubsection{Misaligned Inner Disk}
A precessing misaligned inner disk or change in the inner disk structure is invoked to explain see-saw variability in UX Tau A \citep{espaillat2024}. In that scenario, the inner disk precesses due to the presence of a Jupiter-mass planet, causing variable emission as the viewing angle of the inner disk wall changes with time. Nevertheless, the precession timescale at $R_d$ is $\sim$years, so a precessing warped inner disk still cannot explain the lack of excess emission in the W1/W2 bands two weeks before {\jwst}.
 
\subsubsection{An Arrival of a New Dust Clump}
Lastly, we consider the case where a clump of dust drifted in and had just arrived at the dust sublimation radius at the time of the JWST observations. Considering a baseline flux at the 2005 {\spitzer} level, a dust clump with the same solid angle as considered here will not be detectable as an excess in the W1 and W2 bands if the blackbody temperature is below 800\,K. Taking a conservative estimate of $T_d=1,000$\,K, the dust would be at $R_{d, 1000K}\approx16\,\rstar$, which is $\approx8\,\rstar$ away from the dust sublimation radius. For a dust grain orbiting in a gas disk, the velocity departure from Keplerian velocity due to a headwind is given approximately by \citep{hartmann2009}
\begin{equation}
\delta v \sim \frac{v_K}{2}\left(\frac{c_s}{v_K}\right)^2,
\end{equation}
where $v_K$ is the Keplerian orbital speed and $c_s=\sqrt{2k_BT/\mu m_H}$ is the sound speed. According to \citet{weidenschilling1977}, the maximum possible radial drift velocity is $v_R\lesssim\delta v$, which is $\lesssim0.1\,\kms$ in our case. This means that the dust clump needs $\sim2.8$ yr to travel $8\,\rstar$, far longer than the 2-week timescale seen in our case. Therefore, dust drift cannot explain the observations; any new dust needs to form in situ.
 
\section{Summary and Conclusions} \label{sec:summary}
The disk of {\thestar} is in the last stages of evolution based on its very low mass accretion rate. Here, we present a new observation of {\thestar} with {\jwst} which reveals \added{one of} the most significant flux increase short-ward of $10\,\mu$m observed to date, without any change at longer wavelengths, compared with an archival {\spitzer} observation from 2005. This factor-of-two short-wavelength flux increase is consistent with the presence of warm, optically thick dust near the sublimation radius. A comparison with NEOWISE observations suggests that this increase occurs within two weeks. We argue that the most likely explanation for flux increase and the short timescale is rapid in situ dust formation, potentially from planetesimal collisions near the dust sublimation radius.
 
Multi-epoch observations with {\jwst}, along with high-resolution spectroscopy in the optical to search for traces of refractory elements in the accretion flows \citep[e.g.,][]{micolta2023}, will be required to test this scenario. If confirmed, this will provide strong evidence for the transition from a primordial protoplanetary disk to a debris disk, a scenario expected to occur at the last stages of disk evolution.
 
\begin{acknowledgments}
\noindent
We thank the referee for helpful suggestions that improved the manuscript.
We acknowledge support from {\jwst} grant GO-3983. This work is partially supported by grants for the development of new faculty staff, Ratchadapiseksomphot Fund, Chulalongkorn University. C.P. acknowledges funding from the NSF Graduate Research Fellowship Program under grant No. DGE-1840990.
 
This research has made use of the NASA/IPAC Infrared Science Archive, which is funded by the National Aeronautics and Space Administration and operated by the California Institute of Technology. We also acknowledge the NSTDA Supercomputer Center (ThaiSC) and the National e-Science Infrastructure Consortium for their support of computing facilities. This work was supported by the ThaiSC Voucher Program, Phase I. 
 
This research made use of Astropy,\footnote{http://www.astropy.org} a community-developed core Python package for Astronomy \citep{astropy-collaboration2013,astropy-collaboration2018,astropy-collaboration2022}.
\end{acknowledgments}
 
\facilities{JWST (MIRI), Spitzer (IRS), WIYN (NEID), Magellan:Clay (MIKE), TESS, WISE, NEOWISE, IRSA}

\software{Astropy \citep{astropy-collaboration2013,astropy-collaboration2018,astropy-collaboration2022}, TESSExtractor \citep{serna2021}}
 
All of the {\it JWST} data presented in this article were obtained from the Mikulski Archive for Space Telescopes (MAST) at the Space Telescope Science Institute. The specific observations analyzed can be accessed via \dataset[DOI: 10.17909/qa5z-xj32]{https://doi.org/10.17909/qa5z-xj32}
 
\bibliography{tts}{}
\bibliographystyle{aasjournalv7}

\end{document}